\documentclass[%
 reprint,
 amsmath,amssymb,
 aps,
]{revtex4-1}

\usepackage{graphicx}
\usepackage{dcolumn}
\usepackage{bm}

\begin{document}

\title{Exact analytical solutions for time-dependent Hermitian Hamiltonian systems from static unobservable
	non-Hermitian Hamiltonians}

\author{Andreas Fring}
 \altaffiliation{a.fring@city.ac.uk}
\author{Thomas Frith}%
 \email{thomas.frith@city.ac.uk}
\affiliation{Department of Mathematics, City, University of London,
	Northampton Square, London EC1V 0HB, UK}%

\date{\today}

\begin{abstract}
We propose a procedure to obtain exact analytical solutions to the time-dependent Schr\"{o}dinger
equations involving explicit time-dependent Hermitian Hamitonians from solutions to time-independent non-Hermitian Hamiltonian systems and the time-dependent Dyson relation together with the time-dependent quasi-Hermiticity relation. We illustrate the working of this method for a simple Hermitian Rabi-type model by relating it to a non-Hermitian time-independent system corresponding to the one-site lattice Yang-Lee model.
\begin{description}
\item[03.65.-w,03.65.Aa]
 
\end{description}
\end{abstract}

\maketitle


\section{Introduction}

The time-dependent Schr\"{o}dinger equations (TDSE) is central in the
description of almost all quantum mechanical phenomena. When it involves
explicitly time-depedent Hamiltonians it is usually extremely difficult to solve, so
that only few exact analytical solutions have been found so far. In most
cases one resorts to approximative approaches. For instance, for systems with
quantum potentials and classical electromagnetic fields one relies almost entirely on
perturbative methods either in the weak or strong field 
regime \cite{AC1,Rev1} with only a few available approximative methods that go beyond \cite{BeyondCarla}. 
Schemes that allow to construct exact analytical solutions, such as for instance the method of invariants
proposed by Lewis and Riesenfeld \cite{Lewis69}, are extremely rare and only very few exactly solvable models are known. Thus almost any workable alternative procedure will constitute an advance of the subject area. 

Here we propose a new method that allows in principle to find exact analytical solutions,
but it may also be adapted to a perturbative setting.
Our approach exploits some special solutions of the time-dependent Dyson and
time-dependent quasi-Hermiticity relations \cite{CA,time1,fringmoussa}
in which we take the non-Hermitian Hamiltonian to be time-independent. The
problem of solving the TDSE for a time-dependent Hermitian Hamiltonian $%
h(t)=h^{\dagger }(t)$ is replaced by the much easier one to solve the TDSE
for a time-dependent non-Hermitian Hamiltonian $H\neq H^{\dagger }$ and the
time-dependent Dyson relation for the Dyson map or time-dependent quasi-Hermiticity
relation for the metric operator.

Hence our starting point are the two TDSEs%
\begin{equation}
h(t)\phi (t)=i\hbar \partial _{t}\phi (t),\qquad H\Psi
(t)=i\hbar \partial _{t}\Psi (t),  \label{TS}
\end{equation}%
for which the two wave functions $\phi (t)$ and $\Psi (t)$ are assumed to be
related by a time-dependent invertible operator $\eta (t)$, the time-dependent Dyson map, as%
\begin{equation}
\phi (t)=\eta (t)\Psi (t).  \label{sol}
\end{equation}%
It then follows by direct substitution of (\ref{sol}) into (\ref{TS}) that
the two Hamiltonians are related to each other by the time-dependent Dyson
relation 
\begin{equation}
h(t)=\eta (t)H\eta ^{-1}(t)+i\hbar \partial _{t}\eta (t)\eta ^{-1}(t).
\label{hH}
\end{equation}
Thus computing $\phi (t)$ from the first equation in (\ref{sol}) becomes
equivalent to computing $\Psi (t)$ from the second equation in (\ref{TS})
and $\eta (t)$ from (\ref{hH}). Alternatively we may also compute $\eta (t)$
by solving the quasi-Hermiticity relation  
\begin{equation}
H^{\dagger }\rho (t)-\rho (t)H=i\hbar \partial _{t}\rho (t).  \label{etaH}
\end{equation}%
for the metric operator $\rho (t)$ and subsequently use $\rho (t):=\eta
^{\dagger }(t)\eta (t)$. So despite the fact that the Hamiltonian $H$ is
static we will associate it to a time-dependent metric. We also note, as
previously argued in \cite{fringmoussa},  that because of the presence of
the gauge-like term in (\ref{hH}) the non-Hermitian Hamiltonian $H$ is not
quasi-Hermitian and therefore not observable. Instead the operator 
\begin{equation}
\tilde{H}(t)=\eta ^{-1}(t)h(t)\eta (t)=H+i\hbar \eta ^{-1}(t)\partial
_{t}\eta (t), \label{Hphys}
\end{equation}
is quasi-Hermitian and interpreted as the physical operator that plays the role of the energy in the non-Hermitian system. It does, however, not satisfy the relevant TDSE. One may of course define for $\tilde{H}(t)$ a new time-dependent Schr\"{o}dinger equation $\tilde{H}(t)\tilde{\Psi}(t)=i\hbar \partial _{t}\tilde{\Psi}(t)$, but that would be a new system with different Hilbert space and therefore with different physical content.
Having solved (\ref{sol}), we can subsequently construct an exact form for
the unitary time-evolution operator 
\begin{equation}
u(t,t^{\prime })=T\exp \left[ -i\int\nolimits_{t^{\prime }}^{t}dsh(s)\right]
,  \label{utt}
\end{equation}%
that evolves a state $\phi (t)=u(t,t^{\prime })\phi (t^{\prime })$ from a
time $t^{\prime }$ to $t$ satisfying $h(t)u(t,t^{\prime })=i\hbar \partial _{t}u(t,t^{\prime })$,
$u(t,t^{\prime })u(t^{\prime },t^{\prime \prime })=u(t,t^{\prime \prime
})$ , $u(t,t)=\mathbb{I}$ and preserves by definition the inner product $\left\langle u(t,t^{\prime
})\phi (t^{\prime })\left\vert u(t,t^{\prime })\tilde{\phi}(t^{\prime
})\right\rangle \right. =\left\langle \phi (t)\left\vert \tilde{\phi}%
(t)\right\rangle \right. $. The time-evolution operator $U(t,t^{\prime
})=\eta ^{-1}(t)u(t,t^{\prime })\eta (t^{\prime })$ that evolves states $%
\Psi (t)=U(t,t^{\prime })\Psi (t^{\prime })$ in the non-Hermitian system
from a time $t^{\prime }$ to $t$ is not expected to be unitary in a standard
matrix representation, but it preserves the modified inner product     
\begin{equation}
\left\langle U(t,t^{\prime })\Psi (t^{\prime })\left\vert U(t,t^{\prime })%
\tilde{\Psi}(t^{\prime })\right\rangle \right. _{\rho }=\left\langle \Psi
(t)\left\vert \tilde{\Psi}(t)\right\rangle \right. _{\rho }.
\end{equation}%
and in that sense guarantees unitary time-evolution. 

A priori it is not clear whether the equations above actually admit nontrivial and meaningful solutions
in the way described above. In fact, it was doubted that they make sense at all and were 
interpreted as a no-go theorem for the possibility to have consistent non-Hermitian systems 
with time-dependent metric \cite{time1}. This view was already challenged in \cite{time6} and in \cite{BilaAd,fringmoussa,fring2016non} it was demonstrated that nontrivial solutions exist. Besides mathematical
arguments questioning the solvability of these equations the main physical objection was based on the fact that the Hamiltonian operator $H$, or $H(t)$, that  governs the TDSE is no longer observable. In  \cite{fringmoussa}
it was argued that this is an unnecessary requirement. It is already well accepted that in the non-Hermitian setting many operators, such as for instance the standard position or momentum operator, become mere auxiliary operators that do not correspond to observable quantities. In the time-dependent setting one simply need to add $H(t)$ to that list and interpret $\tilde{H}(t)$ as the observable quantity.

Here we will elaborate on the special type of solutions for which the non-Hermitian Hamiltonian is kept independent of time and study a simple $(2 \times 2)$-matrix Hamiltonian with a periodic time-dependent potential
\begin{equation}
h(t)=-\frac{1}{2}\left[ \omega \mathbb{I+}\frac{2\phi ^{2}}{2+\gamma
	^{2}\sin \left( t\phi \right) -\gamma ^{2}}\sigma _{z}\right] . \label{Rabi}
\end{equation}
where $\sigma _{x,y,z}$ denote the standard Pauli matrices and $\omega,\gamma,\phi \in \mathbb{R}$ are constants constrained as $\phi=\sqrt{1-\gamma^2}$, $\gamma \geq 1$. This Hamiltonian is similar in type
to the Hermitian Rabi model solved in \cite{shirley65} using perturbation theory \cite{salwen55}. A perturbative
treatment of the non-Hermitian $\mathcal{PT}$-symmetric version of this
model was recently considered in \cite{lee15}. Here we are providing a
non-perturbative analytical solution following the procedure outlined above.

\section{A general non Hermitian SU(2)-Hamiltonian}

The non-Hermitian counterpart to $h(t)$ in (\ref{Rabi}) falls into the general class of Hamiltonians built from generators of an SU(2)-Lie algebra represented here by standard Pauli matrices 
\begin{equation}
H(t)=\frac{1}{2}\left[ \kappa _{0}+i\lambda _{0}\right] \mathbb{I}+%
\frac{1}{2}\sum\nolimits_{j=x,y,x}\left[ \kappa _{j}+i\lambda _{j}\right]
\sigma _{j},\quad   \label{HSU2}
\end{equation}%
with $\kappa _{0},\lambda _{0},\kappa _{j},\lambda
_{j}\in \mathbb{R}$. In what follows we will drop the explicit sum and use the
standard sum convention over repeated indices. Trying to solve the
time-dependent quasi-Hermiticity relation (\ref{etaH}) we make the generic
Ansatz%
\begin{equation}
\rho (t)=\alpha (t)\mathbb{I}+\beta _{j}(t)\sigma _{j},\qquad \alpha
(t),\beta _{j}(t)\in \mathbb{R},  \label{rho}
\end{equation}%
for the metric operator. Substituting (\ref{rho}) into (\ref{etaH}) and
reading off the coefficients of the generators then leads to the
constraining first order differental equations%
\begin{eqnarray}
\alpha _{t} &=&-\alpha \lambda _{0}-\vec{\beta}\cdot \vec{\lambda},
\label{c1} \\
\vec{\beta}_{t} &=&\vec{\kappa}\times \vec{\beta}-\lambda _{0}\vec{\beta}%
-\alpha \vec{\lambda},  \label{c2}
\end{eqnarray}%
for the as yet unknown functions $\alpha(t),\beta _{j}(t)$. Demanding the metric operator to be positive definite imposes the additional
constraint%
\begin{equation}
\det \rho =\alpha ^{2}-\vec{\beta}\cdot \vec{\beta}>0.  \label{c3}
\end{equation}%
Next we will solve the constraints (\ref{c1})-(\ref{c3}).

\subsection{Time-independent Hamiltonian and time-independent metric}

At first we consider the simplest scenario that is obtained when we just
reduce the equations to the standard time-independent scenario, see \cite{Benderrev,Alirev} for reviews. In this case
(\ref{c1})-(\ref{c2}) simplify to%
\begin{equation}
\vec{\beta}\cdot \vec{\lambda}=-\alpha \lambda _{0},\quad \text{and\quad }%
\vec{\kappa}\times \vec{\beta}=\lambda _{0}\vec{\beta}+\alpha \vec{\lambda},
\end{equation}%
which when taking $\lambda _{0}=0$ is easily solved by%
\begin{equation}
\vec{\beta}=\frac{\alpha }{\left\vert \vec{\kappa}\right\vert ^{2}}\vec{%
	\lambda}\times \vec{\kappa}+\nu \vec{\kappa},\qquad \nu \in \mathbb{R}.
\end{equation}%
This solution was also reported in \cite{BilaAd} and only serves here as a
benchmark when taking the limit to the time-independent case. Using the
parameterization (\ref{rho}) then reproduces solutions previously obtained
for the Hamiltonians falling into the class reported in (\ref{HSU2}) for the
time-independent scenario, see for instance \cite{Bender:2002vv} for an example.

\subsection{Time-independent Hamiltonian and time-dependent metric}

\label{sectiHtdrheo}

Next we allow $\partial _{t}\rho $ to be non-vanishing so that $H$ is no
longer quasi-Hermitian because (\ref{etaH}) has a nonvanishing right hand
side. Guided by the solution in the previous section we keep $\lambda _{0}=0$ and substitute the Ansatz%
\begin{equation}
\vec{\beta}(t)=\zeta _{1}(t)\vec{\kappa}+\zeta _{2}(t)\vec{\lambda}+\zeta
_{3}(t)\vec{\kappa}\times \vec{\lambda}
\end{equation}%
into (\ref{rho}) and (\ref{etaH}) and thus obtaining a set of simple first order coupled
differential equations 
\begin{equation}
\partial _{t}\zeta _{1}=\zeta _{3}\vec{\kappa}\cdot \vec{\lambda},
\quad \partial _{t}\zeta _{2}=-\alpha -\zeta _{3}\left\vert \vec{\kappa}%
\right\vert ^{2},\quad \partial _{t}\zeta _{3}=\zeta _{2},
\end{equation}%
as constraints. Assuming that $\vec{\kappa}\cdot \vec{\lambda}=0$ the
general solutions to these equations are easily obtained as%
\begin{eqnarray}
\zeta _{1}(t) &=&c_{4}, \\
\zeta _{2}(t) &=&c_{1}\sin (\phi t)+c_{2}\cos (\phi t), \\
\zeta _{3}(t) &=&-\frac{c_{1}}{\phi }\cos (\phi t)+\frac{c_{2}}{\phi }\sin
(\phi t)+c_{3}, \\
\alpha (t) &=&\left( \frac{c_{1}}{\phi }\left\vert 
\vec{\kappa}\right\vert ^{2}-c_{1}\phi \right) \cos (\phi t)\\
&& +\left( c_{2}\phi -\frac{c_{2}}{\phi }\left\vert \vec{\kappa}%
\right\vert ^{2}\right) \sin (\phi t)-c_{3}\left\vert 
  \vec{\kappa}\right\vert ^{2} \qquad \qquad \qquad
\end{eqnarray}%
with $\phi =%
\sqrt{\vert \vec{\kappa}\vert ^{2}-\vert \vec{\lambda}%
	\vert ^{2}}$ and $c_{1},\ldots ,c_{4}\in \mathbb{R}$ being arbitrary constants. 

Having obtained $\rho(t)$ we may easily compute $\eta(t)$, but in order to carry out the second step in our procedure, that is solving the TDSE, we need to be more specific. Let us therefore study a concrete model that falls into the general class of Hamiltonians treated in this section.

\section{The one-site lattice Yang-Lee model}

We consider an Ising quantum spin chain in the presence of a magnetic field
in the $z$-direction and a longitudinal imaginary field in the $x$-direction 
\cite{chainOla} that has been identified as the discretised lattice version
of the Yang-Lee model \cite{gehlen1}, described by the non-Hermitian
Hamiltonian 
\begin{equation}
H_{N}=-\frac{1}{2}\sum\nolimits_{j=1}^{N}(\sigma _{j}^{z}+\lambda \sigma
_{j}^{x}\sigma _{j+1}^{x}+i\kappa (t)\sigma _{j}^{x}),  \label{H}
\end{equation}
with $\lambda,\kappa \in \mathbb{C}$. In \cite{fringmoussa} it was demonstrated that when taking $N=1$ the
time-dependent quasi-Hermiticity relation admits nontrivial solutions. In
this case the non-Hermitian Hamiltonian just reduces to a particular example of (\ref{HSU2}) 
\begin{equation}
H_{1}=-\frac{1}{2}\left[ \lambda 
\mathbb{I}+\sigma _{z}+i\kappa \sigma _{x}\right] .
\end{equation}%
We will now solve the time-dependent Dyson relation\ (\ref{hH}) together with the
time-dependent quasi-Hermiticity relation (\ref{etaH}) and the TDSE for $H_{1}$ in more detail. 

\subsection{Time-independent Hamiltonian and time-independent metric}

Specifying the quantities in section II A as $\lambda =\omega \equiv $ const and $\kappa =\gamma \equiv $
const we identify $\kappa _{0}=-\omega $, $\lambda _{0}=0$, $\vec{\kappa}%
=(0,0,-1)$, $\vec{\lambda}=(-\gamma ,0,0)$, so
that%
\begin{equation}
\vec{\beta}=-\alpha \gamma \vec{e}_{y}-\nu \vec{e}_{z},\quad \rho =\alpha 
\mathbb{I}-\alpha \gamma \sigma _{y}-\nu \sigma _{z},
\end{equation}%
with $\det \rho =\alpha^{2}(1-\gamma ^{2})-\nu ^{2}>0$ and $\vec{e}_{i}$ denoting unit vector in the direction $i=x,y,z$. As expected, the metric ceases to be positive definite when the eigenvalues 
\begin{equation}
E_{\pm }=\frac{1}{2}\left( -\omega \pm \phi \right)  \label{EPM}
\end{equation}%
of $H$ become complex conjugate, that is when $\gamma >1$. Taking now $\nu
=0 $ for simplicity, we compute the Dyson map $\eta $ as the square root of
the metric $\rho $ as%
\begin{equation}
\eta =\sqrt{\rho }=\frac{\sqrt{\alpha }}{2}\left[ \left( \phi_{-}+%
\phi_{+}\right) \mathbb{I+}\left( \phi_{-}-\phi_{+}
\right) \sigma _{y}\right] .
\end{equation}
with $\phi_{\pm} =   \sqrt{1 \pm \gamma } $. Assuming $\eta$ to be Hermitian, it is computed from $\rho $ simply
by taking the square root in the standard way by diagonalizing it first
as $\rho =UDU^{-1}$ and subsequently computing $\sqrt{\rho }=UD^{1/2}U^{-1}$. Using this expression in (\ref{hH}) leads to the isospectral Hermitian counterpart
\begin{equation}
h=-\frac{1}{2}\left( \omega \mathbb{I}+\phi \sigma _{z}\right) .
\end{equation}%
It is of course well known how to obtain these type of relations in the
time-independent case, but the expressions obtained here serve as benchmarks
for the time-dependent case.

\subsection{Time-independent Hamiltonian and time-dependent metric}

Switching on the time-dependence we solve first the TDSE. Since $H_1$ is time-independent 
this is easily achieved by expanding the solution in terms of the energy eigenstates $E_{\pm }$ 
in (\ref{EPM}) as $\Psi _{\pm }(t)=c_{\pm }e^{-iE_{\pm }t}$ with some constants $c_{\pm }$.
Substitution of this Ansatz into (\ref{TS}) then yields the solutions to (\ref{TS}) with a suitable normalization
\begin{equation}
\Psi _{\pm }(t)=\frac{\sqrt{\gamma }}{\sqrt{2}\phi \sqrt{1\pm \phi }} \left( 
\begin{array}{c}
\gamma \\ 
i(1\pm \phi )%
\end{array}%
\right) e^{-iE_{\pm }t}. \label{solphi}
\end{equation}%
Next we need to solve (\ref{hH}) and (\ref{etaH}), for $\eta(t)$ and $\rho(t)$, respectively. Keeping at first all integration constants generic we obtain from (\ref{c3}) 
\begin{equation}
\det [\rho (t)]=c_{3}^{2}-c_{4}^{2}-\gamma
^{2}(c_{1}^{2}+c_{2}^{2}+c_{3}^{2})>0.
\end{equation}%
Thus it is vital to maintain $c_{3}\neq 0$. Considering the solution in section %
\ref{sectiHtdrheo}, we first notice that $\partial _{t}\rho =0$ leads to $%
c_{1}=c_{2}=0$, so that we recover the time-independent scenario in this
case. A nontrivial convenient choice is for instance $c_{1}=0$, $c_{2}=-\phi /\gamma $, $%
c_{3}=-1/\gamma $, $c_{4}=0$, leading to the time-dependent metric 
\begin{equation}
\rho (t)=\left[ \frac{1}{\gamma}+\gamma \sin (\phi t)\right] \mathbb{I}+\phi
\cos (\phi t)\sigma _{x}-\left[ 1+\sin (\phi t)\right] \sigma _{y}. \label{solrho}
\end{equation}%
Taking the square root, similarly as in the previous section, then yields
the time-dependent Dyson map%
\begin{eqnarray}
\eta (t)&=& \frac{1}{2}\left[ p_{+}(t)+p_{-}(t)\right] \mathbb{I}  \label{soleta} \\
&& +  \frac{ p_{+}(t)-p_{-}(t)}{%
	2\left\vert p_{0}(t)\right\vert } \left[ {Im}\left[ p_{0}(t)\right]\sigma _{x} -{Re}\left[ p_{0}(t)\right]\sigma _{y} \right] \nonumber ,
\end{eqnarray}
where we abbreviated the functions
\begin{eqnarray}
p_{\pm }(t)&=&\sqrt{\gamma ^{-1}+\gamma \sin (\phi t)\pm \left\vert 
	p_{0}(t)\right\vert }, \\
 p_{0}(t)&=&1+\sin (\phi t)+i\phi \cos (\phi t).
\end{eqnarray}%
Using this expression for $\eta (t)$ in (\ref{hH}) produces the Hermitian
time-dependent Hamiltonian $h(t)$ in (\ref{Rabi}). 

We have now obtained explicit analytical solutions for all time-dependent wave functions. Next we verify that
they yield meaningful expectation values. Using (\ref{sol}) together with our solutions (\ref{solphi}), (\ref{solrho}) and (\ref{soleta}), we compute
\begin{eqnarray}
\left\langle \Psi _{\pm }(t)\right. \left\vert \rho (t)\Psi _{\pm
}(t)\right\rangle &=&\left\langle \phi _{\pm }(t)\right. \left\vert \phi
_{\pm }(t)\right\rangle =1,\quad \\
\left\langle \Psi _{\mp }(t)\right. \left\vert \rho (t)\Psi _{\pm
}(t)\right\rangle &=&\left\langle \phi _{\mp }(t)\right. \left\vert \phi
_{\pm }(t)\right\rangle =\pm i\gamma .
\end{eqnarray}
These states were not expected to be orthonormal, but we can use them to easily find an orthonormal basis. A useful and natural basis is
\begin{eqnarray}
\phi _{1}(t_{0}) &=&\binom{1}{0}=c_{+}\left\vert \phi
_{-}(t_{0})\right\rangle +c_{-}\left\vert \phi _{+}(t_{0})\right\rangle
\label{f1} \\
\phi _{2}(t_{0}) &=&\binom{0}{1}=c_{-}\left\vert \phi
_{-}(t_{0})\right\rangle -c_{+}\left\vert \phi _{+}(t_{0})\right\rangle
\label{f2}
\end{eqnarray}%
with%
\begin{equation}
t_{0}=-\frac{\pi }{2\phi },\qquad c_{\pm }=\frac{1}{\sqrt{2}\phi ^{2}}e^{%
	\frac{i\pi }{4}\left( \frac{\omega }{\phi }\pm 1\right) }\left( \phi_\pm - \gamma \phi_\mp  \right) .
\end{equation}%
Using these states and $t_0$ as initial time, the time-evolution operator is easily extracted from the explicit form of $\phi(t)$ as
\begin{equation}
u(t,t_{0})=\left( 
\begin{array}{cc}
e^{i \theta(t)} & 0 \\ 
0 & e^{\frac{i\pi }{2}\left( \frac{\omega }{\phi }+\frac{2t\omega }{\pi }%
	\right) -i \theta(t) }
\end{array}%
\right)  \label{solu}
\end{equation}%
with%
\begin{equation}
\theta(t)= \frac{\pi}{4} + \frac{\omega}{2}(t-t_0)+ \arctan \left[ \frac{(1-\phi )^2+\gamma \tan
	\left( \frac{t\phi }{2}\right) }{\gamma+(1-\phi )^2\tan \left( \frac{t\phi 
	}{2}\right) }\right]  .
\end{equation}
One may verify that $u(t,t_{0})$ indeed satisfies the TDSE. It is now also straightforward to compute the
time-evolution operator for the non-Hermitian system $U(t,t_0)=\eta ^{-1}(t)u(t,t_0)\eta (t_0)$ using (\ref{soleta})
and (\ref{solu}).

In order to sustain our claim that the Hamiltonian $\tilde{H}(t)$ defined in (\ref{Hphys}) represents the
energy in the non-Hermitian system we compute the energy expectation values 
\begin{eqnarray}
E_{\pm }(t)&=&\left\langle \Psi _{\pm }(t)\right. \left\vert \tilde{H}(t)\rho
(t)\Psi _{\pm }(t)\right\rangle \\ 
&=&\left\langle \phi _{\pm }(t)\right.
\left\vert h(t)\phi _{\pm }(t)\right\rangle \\
&=&\pm \frac{\phi ^{3}}{2+\gamma
	^{2}\sin \left( t\phi \right) -\gamma ^{2}}-\frac{\omega }{2}  \label{eexp}
\end{eqnarray}%
oscillating with Rabi-frequency $\phi =(E_{+}-E_{-})$ between the values $E_\pm(t_0)=E_\pm$ and $E_\pm(-t_0)=(\pm \phi ^{3}-\omega )/2$. Thus $h(t)$ the Hermitian side corresponds to $\tilde{H}(t$ on the non-Hermitian side. 

\medskip

\section{Conclusions}

We have demonstrated that the problem of solving the TDSE involving an explicitly time-dependent Hermitian
Hamiltonian (\ref{TS}) can be replaced with a two step procedure consisting of first solving the TDSE for a
time-independent non-Hermitian Hamiltonian (\ref{TS}) and second solving the time-dependent Dyson relation (\ref{hH}) together with the time-dependent quasi-Hermiticity relation (\ref{etaH}). For the simple model presented here it transpires that the equations in our two-step procedure are indeed much easier to solve than the original TDSE. Here we have presented a derivation with $H$ as starting point, but of course all steps are reversible and one may also take $h(t)$ to commence with. 

As a by-product we have also obtained further evidence for the solvability of the equations (\ref{hH}) and (\ref{etaH}), as already observed in \cite{fringmoussa,fring2016non}, but in addition we also showed here that the solutions obtained constitute meaningful wavefunctions and produce physical expectation values.

Clearly our approach can also be adapted to a perturbative treatment. Just as in the time-independent setting, where the Dyson map is often only known perturbatively  \cite{Benderrev,Alirev}, this limitation is likely to carry over to the time-dependent scenario. So a perturbative series would be in a parameter related to $\eta$ rather than one occurring in the model itself. 

It will be very interesting to investigate the viability of this approach further for more complicated systems of higher rank matrix type, but especially for Hamiltonians related to infinite dimensional Hilbert spaces. More investigations are also desirable for the situation in which the non-Hermitian Hamiltonian in (\ref{TS}) are explicitly time-dependent.

\bigskip \noindent \textbf{Acknowledgments:} TF is supported by a City,
University of London Research Fellowship. AF thanks Carla Figueira de Morisson Faria
for useful discussions and comments. 


%

\end{document}